\documentclass[twocolumn,showpacs,preprintnumbers,amsmath,amssymb]{revtex4}
\usepackage{graphicx}
\usepackage{dcolumn}
\usepackage{bm}

\newcommand{\kslash}{k \hskip -0.5em /}
\newcommand{\sss}{\scriptscriptstyle s}

\newcommand{\bgi}{\begin{itemize}}
\newcommand{\eni}{\end{itemize}}

\newcommand{\bb}{\begin{thebibliography}}
\newcommand{\eb}{\end{thebibliography}}

\newcommand{\nn}{\nonumber \\}
\newcommand{\bwt}{\begin{widetext}}
\newcommand{\ewt}{\end{widetext}}
\newcommand{\bea}{\begin{eqnarray}}
\newcommand{\ba}{\begin{array}}
\newcommand{\ea}{\end{array}}
\newcommand{\eea}{\end{eqnarray}}
\newcommand{\be}{\begin{equation}}
\newcommand{\ee}{\end{equation}}

\newcommand{\bit}[1]{\bibitem{#1}}

\newcommand{\nnn}{\nonumber}

\newfont{\fib}{cmfi10 at 10pt}

\newcommand{\A}{{\cal A}}

\newcommand{\eg}{{\it e.g.}\ }

\begin{document}

\preprint{BK-PSU:HEP 01-03}

\title{Novel Transversity Properties in Semi-Inclusive Deep Inelastic Scattering}

\author{
Leonard P.  Gamberg,$^{a\star}$
Gary R. Goldstein,$^{b,\dagger}$
Karo A. Oganessyan,$^{c,d,\diamond}$} 
\address{
$^a$Division of Science,
Penn State-Berks Lehigh Valley College, 
Reading, PA 19610, USA \\
$^b$ Department of Physics and Astronomy, Tufts University,
           Medford, MA 02155, USA \\
$^c$ INFN-Laboratori Nazionali di Frascati I-00044 Frascati, 
via Enrico Fermi 40, Italy\\
$^d$ DESY, Deutsches Elektronen Synchrotron 
Notkestrasse 85, 22603 Hamburg, Germany \\
$^\star${\sf (lpg10@psu.edu)}\\
$^\dagger${\sf (gary.goldstein@tufts.edu)}\\
$^\diamond${\sf (kogan@mail.desy.de)}
}
\date{\today}
\begin{abstract}
The $T$-odd distribution functions contributing to transversity
properties of the nucleon  and their role in fueling
nontrivial contributions to 
azimuthal asymmetries in  semi-inclusive deep inelastic scattering
are investigated. 
We use a dynamical model to evaluate these quantities in
terms of HERMES kinematics.  
\end{abstract}

\maketitle

%
\section{\label{intro} Introduction}
\vskip-0.25cm
{\it Transversity}, as combinations of helicity states
for the moving nucleon, is a variable introduced 
originally~\cite{gold_morv} to reveal an underlying simplicity in
nucleon--nucleon spin dependent scattering amplitudes.
The  connection with the spin structure of the nucleon
in terms of the chiral odd distribution $h_1(x)$
and its first moment, the tensor charge, 
emerged in the analyses of  hard processes~\cite{jaffe91}.  
These  analyses revealed that to leading order in inverse powers of $Q^2$,
transversity is strongly suppressed in deep inelastic scattering
and the non-conservation of the tensor charge 
makes predicting transversity fairly difficult.
Until recently, our theoretical knowledge of transversity appeared 
limited to suggestive
bounds placed on the leading twist quark 
distributions  in the form of the  inequality of 
Soffer~\cite{soffer95}. In principle, however
transversity can be probed when  at least two hadrons are
present (as was known earlier \eg Drell-Yan~\cite{ralston79}) 
or in semi-inclusive deep inelastic scattering (SIDIS)~\cite{cnpb93}.
These  analyses inspired the first searches
for transversity properties of the nucleon~\cite{RHIC,HERMES}. 
Further, while  probing transversity through
exclusive single meson production 
has met with negative results~\cite{cds}, it
has been suggested that the chiral odd off forward distribution
can be probed when two vector mesons are observed~\cite{pire}. 
Additionally, when quark distributions are enriched with $k_\perp$ 
dependence~\cite{kotz0,mulders2}, while also
allowing for  time reversal odd ($T$-odd) 
behavior~\cite{ANSL,boer,cplb},   
novel transversity properties of quarks in hadrons
can be measured via asymmetries in semi-inclusive
and polarized spin processes. 
For example, the  distribution $f_{1T}^\perp(x,k_\perp)$, representing 
the number density of {\em unpolarized} quarks in transversely polarized
nucleons, may be  entering the 
recent measurements of single spin asymmetries (SSAs) at
HERMES~\cite{HERMES} and at 
RHIC~\cite{bland} in semi-inclusive pion electroproduction.
Alternatively $h_1^\perp(x,k_\perp)$, which 
describes the transfer of  transversity to quarks inside unpolarized
hadrons, may enter transverse momentum dependent 
asymmetries~\cite{boer,cplb,gold_gamb,gamb_gold_ogan,bbh}.  
Beyond the $T$-odd properties, the existence of these distributions
is a signal of the {\em essential} role 
played by the intrinsic transverse quark
momentum and the corresponding angular momentum of quarks inside
the target and fragmenting hadrons  in these hard scattering processes.  
In this paper we analyze the $T$-odd transversity properties of quarks
in hadrons which emerge in SIDIS. We apply these results to 
predict $\cos2\phi$~\cite{boer} and Sivers~\cite{sivers}  asymmetries 
 in terms of HERMES kinematics.

From the standpoint of quark-target helicity flip 
amplitudes, what has emerged from analyzing transversity 
in  inclusive, semi-inclusive, and exclusive processes
is that angular momentum conservation requires
that helicity changes are
accompanied by transferring 1 or 2 units of 
orbital angular momentum, {\em  highlighting}
the essential role of intrinsic $k_\perp$ and
orbital angular momentum in determining transversity. 
The interdependence of transversity on
quark {\em orbital} angular momentum and $k_\perp$
is a general property.
This behavior was revealed 
in studying the~\cite{gamb_gold} vertex function 
associated with the tensor charge and in generalized
meson production amplitudes in exchange models. 
In these cases, angular momentum conservation results in  the transfer of
orbital angular momentum $\ell=1$ carried by the  dominant
$J^{PC}=1^{+-}$ mesons  to compensate for the 
non-conservation of helicity across the vertex.  
The corresponding {\em conjugate} dependence on  
powers of the intrinsic quark
momentum is determined by  these tensor couplings which 
involve helicity flips associated with kinematic factors
of $3$-momentum transfer, as required by angular momentum conservation.
This $k_\perp$ dependence can be understood on fairly general grounds
from the kinematics of the exchange picture
in exclusive pseudoscalar meson photoproduction
and points to the fundamental role of rescattering as the source for
nontrivial leading twist SSAs associated with transversity.

For large $s$ and relatively small momentum transfer $t$
combinations of
the four helicity amplitudes involve definite parity 
exchanges~\cite{gold_owens}. The four independent helicity amplitudes 
can have the minimum kinematically allowed powers,
\bea
&& f_1 = f_{1+,0+} \propto k_\perp^1 , \quad
f_2 = f_{1+,0-} \propto k_\perp^0 , 
\nn &&
\quad
f_3 = f_{1-,0+} \propto k_\perp^2 , \quad
f_4 = f_{1-,0-} \propto k_\perp^1. 
\eea
However, in single hadron exchange (or Regge pole exchange)
parity conservation requires {\small $f_1 =\pm f_4 \quad {\rm and} \quad f_2 =\mp f_3$} for even/odd parity exchanges. These pair relations,
along with a single hadron exchange model,
force $f_2$ to behave like  $f_3$ for small
$t$. This introduces the $k_\perp^2$ factor into $f_2$. However for a
non-zero polarized target asymmetry  to arise there must be
interference between single helicity flip and non-flip and/or double
flip amplitudes. Thus this asymmetry must arise from rescattering
corrections (or Regge cuts or eikonalization or loop corrections) 
to single hadron exchanges. That is, one of the amplitudes in
\be
P_y = \frac{2 Im (f_1^*f_3 - f_4^*f_2)}
{\sum_{j=1...4}|f_j|^2}
\ee
must acquire a different phase.
In fact rescattering  reinstates $f_2 \propto k_\perp^0$ 
by integrating over loop $k_\perp$, which
effectively introduces a $\left< k_\perp^2 \right>$
factor~\cite{gold_owens}. This is true for the {\em inclusive process} 
as well, where only one final hadron is measured; a relative phase
in a helicity flip three body amplitude is required~\cite{gold_owens2}.

\section{T-Odd Distributions 
in Semi-Inclusive Reactions }

Recently  rescattering was considered as a mechanism 
for SSAs in pion electroproduction from transversely polarized
nucleons.  Using the QCD motivated quark-diquark model of the 
nucleon~\cite{rodriq,bhs}, the $T$-odd distribution function,
$f_{1T}^\perp(x,k_\perp)$ and  the corresponding analyzing power for 
the azimuthal asymmetry
in the fragmenting hadron's momentum and spin distributions
resulted in a leading twist nonzero Sivers~\cite{sivers} 
asymmetry~\cite{ji,cplb}.  Using the approach in Ref.~\cite{ji},
 we  investigated the 
rescattering in terms of final state interactions (FSI) 
to the $T$-odd function $h_1^\perp(x)$ 
and corresponding azimuthal asymmetry in 
SIDIS~\cite{gold_gamb,gamb_gold_ogan}. The  asymmetry 
involves the convolution with the $T$-odd fragmentation
function, $h_1^\perp(x)\star H_1^\perp(z)$~\cite{boer}. 
As mentioned in the introduction 
$h_1^\perp(x,k_\perp)$ is complimentary to the Sivers function and
is of great interest theoretically,  since it vanishes at tree 
level, and experimentally, since its determination does not 
involve polarized nucleons~\cite{boer,cplb,gold_gamb,gamb_gold_ogan,bbh}.  

The $T$-odd distributions are readily defined from the 
transverse momentum dependent quark distributions~\cite{col82,cplb} 
where the well known identities for manipulating the limits of
an ordered exponential lead to the expression
\bea
\Phi^{[\Gamma]}(x,k_\perp)&=&{\frac{1} {2}}\sum_n
\int {\frac{d\xi^- d^2\xi_\perp }
  {(2\pi)^3}} e^{-i(\xi^- k^+-\vec{\xi}_\perp \vec{k}_\perp)} 
\nn  && \hspace{-2.7cm}
\times
\langle P|\overline{\psi}(\xi^-,\xi_\perp){\cal G}^{\dagger}(\infty,\xi)
\big|n\rangle\Gamma\langle n\big|
{\cal G}(\infty,0)\psi(0)|P\rangle\vert_{{\scriptscriptstyle{\xi^+=0}}}
\label{link}
\eea
and the path ordered exponential is 
\bea
{\cal G}(\infty,\xi)={\cal P}
\exp{\left(-ig\int_{\xi^-}^\infty d\xi^- A^+(\xi)\right)},
\nnn
\eea
and $\{\big|n\rangle \}$ are a complete set of states.
While the path ordered light-cone link operator is necessary to maintain 
gauge invariance and appears to respect factorization~\cite{cplb,ji,ji2} 
when transverse momentum distributions are considered, 
in non-singular gauges~\cite{ji,ji2}, it also provides a mechanism
to generate  interactions between an eikonalized struck
quark and the remaining target.
These final state interactions in turn gives rise to leading 
twist contribution to the distribution functions that fuel the 
novel SSAs that have been reported in the 
literature~\cite{bhs,cplb,ji,ji2,gold_gamb,bbh,gamb_gold_ogan}.

The Feynman rules for eikonal lines and  vertices  were  derived
sometime ago~\cite{col82,cplb} and applied to the $T$-odd Sivers 
function~\cite{ji} and $h_1^\perp$~\cite{gold_gamb,gamb_gold_ogan} recently.  
They are obtained by expanding the interactions in
the path ordered gauge link operator in Eq.~(\ref{link}).
With the tree level contribution vanishing, 
the leading order  one loop contribution of final state interactions
to the $T$-odd transverse quark distribution function comes from the
first non-trivial term in the expansion. Modeling the remaining 
target in the quark-diquark model~\cite{rodriq,ji}, $\Phi^{[\Gamma]}$
takes the form
\bea
&&\Phi^{[\Gamma]}(x,k_\perp)={\frac{1} {2}}\sum_n
\int {\frac{d\xi^- d^2\xi_\perp }
  {(2\pi)^3}} e^{-i(\xi^- k^+-\vec{\xi}_\perp \vec{k}_\perp)} 
\nn && 
\hspace{.3cm}
\langle P|\overline{\psi}(\xi^-,\xi_\perp)|n\rangle
\langle n|\left(-ie_1\int^\infty_0 A^+(\xi^-,0) d\xi^-
   \right)
\nn && \hspace{2cm} \times\, 
\Gamma\psi(0,0_\perp)|P\rangle\vert_{\scriptscriptstyle{\xi^+=0}}  
\quad +\quad {\rm h. c.}\, ,
\label{phi}
\eea
where $e_1$ is the charge  of the struck quark and now $\{\big|n\rangle \}$  
represents intermediate scalar di-quark spectator states.
The quark-nucleon-spectator model used in previous rescattering 
calculations assumed a point-like nucleon-quark-diquark vertex, which 
leads to logarithmically divergent, $x$-dependent distributions.  
Yet we know there is a distribution of intrinsic transverse 
momenta among the 
constituents of the nucleon, as Drell-Yan processes show~\cite{ellis}. 
To account for this fact and to also address the $\log$ 
divergence~\cite{bhs,ji,gold_gamb,bbh,gamb_gold_ogan} we
assume the transverse momentum dependence of the quark-nucleon-spectator 
vertex can be approximated by a Gaussian distribution in 
$k_\perp^2$~\cite{gamb_gold}
\be
\langle n\big| \psi(0)\big| P\rangle= 
\left(\frac{i}{\kslash - m}\right)
\frac{g_{\sss}}{{\scriptstyle< k_\perp^2>}\pi} 
e^{-\frac{k^2_{\perp}}{<k_\perp^2>}}\, U(P,S)\ ,
\ee
where $g_{\sss}$ (defined henceforth as $g$) is the scalar diquark coupling~\cite{rodriq,bhs}, 
$k$ is the momentum of the quark in the target proton,
$k_\perp$ and $<k_\perp^2>$ are the  intrinsic and average intrinsic
transverse momentum respectively, and   $U(P,S)$ 
is the nucleon spinor. The quark propagator comes from 
the untruncated quark line. Going to momentum space, 
performing the loop integration,
and finally projecting the unpolarized
piece from $\Phi^{\scriptscriptstyle [i\sigma^{\perp +}\gamma_5]}$    
results in the
leading twist, $T$-odd, unpolarized contribution~\cite{new}
\bea
\Phi^{\scriptscriptstyle [\sigma^{\perp +}\gamma_5]}_{\scriptscriptstyle 
[h_1^\perp]}\hspace{-.15cm}&=&\hspace{-.15cm} 
\frac{\varepsilon_{
\scriptscriptstyle 
+-\perp j}k_{\scriptscriptstyle \perp j}}{M}h_1^\perp(x,k_\perp)
\nn &=&\frac{e_1e_2g^2}{2(2\pi)^4}\frac{b^2}{\pi^2}
\frac{(m+xM)(1-x)}{\Lambda(k^2_\perp)}
\frac{\varepsilon_{\scriptscriptstyle +-\perp j}
k_{\scriptscriptstyle \perp j}}{k_\perp^2}
\nn &&\times
e^{-b\left(k^2_\perp- \Lambda(0)\right)}
\hspace{-.15cm}\left[\Gamma(0,b\Lambda(0))\hspace{-.10cm}-\hspace{-.10cm}
\Gamma(0,b\Lambda(k^2_\perp))\right] .
\nn
\eea
Here, $e_2$ is the gluon-scalar diquark coupling, and 
$\Lambda(k^2_\perp)=k_\perp^2 +(1-x)m^2 +x\lambda^2  -x(1-x)M^2$, where 
$M$, $m$, and $\lambda$ are the nucleon, quark, and diquark masses 
respectively.  Also, $b=\frac{1}{<k_\perp^2>}$, where  $<k_\perp^2>$ 
is fixed below.  As a check on our approach, letting   $b$  go 
to zero which is equivalent to letting  $<k_\perp^2>\rightarrow \infty$
and expanding the incomplete gamma function $\Gamma(0,z)$ in powers of 
 $z=b\Lambda$,
we obtain the log divergent  result~\cite{gold_gamb,gamb_gold_ogan,bbh}.   
The average $k^2_{\perp}$ is a regulating scale 
which we fit to the expression for the integrated unpolarized 
structure function
\bea
f(x)&=&\frac{g^2}{(2\pi)^2}\frac{b^2}{\pi^2}\left(1-x\right)
\nn &&
\times\Bigg\{\frac{\left(m+xM\right)^2-\Lambda(0)}{\Lambda(0)}
\nn &&\hspace{-1cm}
-\left[2b\left(\left(m+xM\right)^2-\Lambda(0)\right)-1\right]
e^{2b\Lambda(0)}\Gamma(0,2b\Lambda(0))\Bigg\}\, ,
\nn
\eea
which multiplied by $x$ at $<k_\perp^2> = {(0.4)}^2$ GeV$^2$ 
is in  good agreement with the valence
distribution of Ref.~\cite{GRV}.
 
The $T$-odd distribution is  leading twist and IR 
finite and thus provides a phenomenological basis 
from which to model the fragmentation process,  
which along with the quark fragmentation function $\Delta(p)$ 
 enters the hadronic tensor to leading 
order in $1/Q^2$~\cite{mulders2}  
\bwt
\bea
M{\cal W}^{\mu\nu}(P,P_h,q)=\int d^4\, k d^4\, p\delta^4(k+q-p)
{\rm Tr}\left(\gamma^\mu\Phi(k)\gamma^\nu\Delta(p)\right)
+\left(
\begin{array}{ccc} 
q & \leftrightarrow  & -q \\ 
\mu & \leftrightarrow  & \nu
\end{array}
\right).
\eea
\ewt
Here $k$ and $p$ are the quark scattering and 
fragmenting momenta.
We consider this to be a reasonable phenomenological model, which avoids 
the log divergence involved in integrating over $k_\perp$, 
while introducing an average transverse momentum determined from spin 
averaged scattering.  Additionally, this 
form factor approach is compatible with the
parameterization of the fragmentation functions employed 
in~\cite{mulders2,kotz,OABD}
to set the Gaussian width for the fragmentation function.
It is worth noting that the functions $h^{\perp}_1$ and 
$f^{\perp}_{1T}$ are equal (up to a sign) in the quark-scalar diquark 
model~\cite{gold_gamb,bbh}.  
In the model calculation, the Sivers
function is obtained by letting $\Gamma=\gamma^+$ in Eq.~(\ref{phi}) 
with the target polarized in the transverse direction~\cite{bhs,ji,bbh}.  
This supports 
the suggestion that the spin-independent $\cos2\phi$ 
and single-spin $\sin(\phi-\phi_S)$ Sivers asymmetries are 
closely related in hard scattering processes.

\section{Asymmetries}

We discuss the explicit results and numerical evaluation of the
spin-independent double $T$-odd $\cos 2\phi$ and single 
transverse-spin $\sin (\phi-\phi_s)$ asymmetries for
$\pi^+$ production in SIDIS.
\begin{figure*}
\includegraphics[height=6.0 cm]{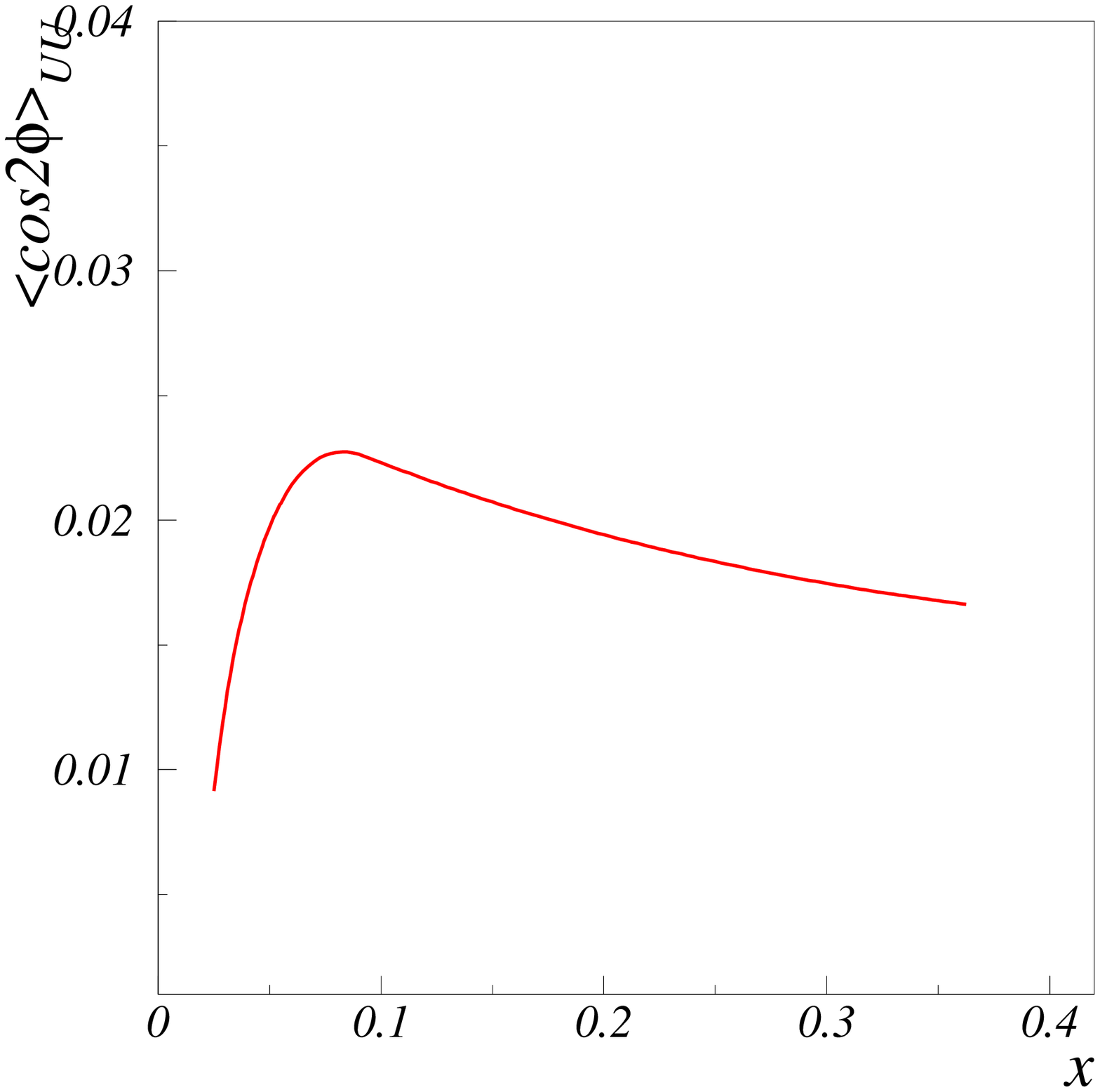}
\includegraphics[height=6.0 cm]{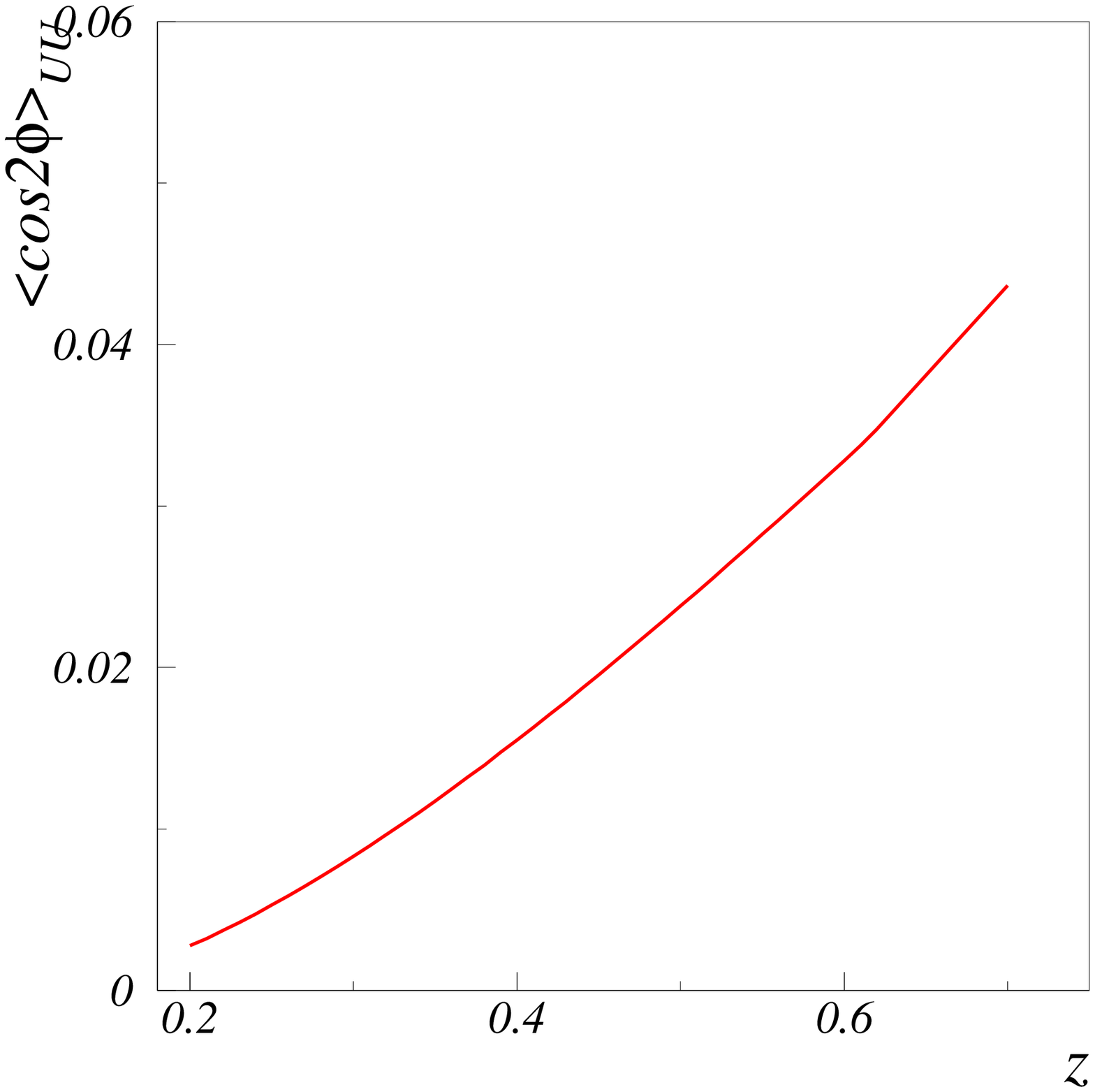}
\caption{\label{A}Left Panel: The \protect{${\langle \cos2\phi 
\rangle}_{\scriptscriptstyle UU}$} asymmetry for 
\protect{$\pi^+$} production 
as a function of \protect{$x$}. 
Right Panel:
The 
\protect{${\langle \cos2\phi \rangle}_{\scriptscriptstyle UU}$} asymmetry for 
\protect{$\pi^+$} production 
as a function of \protect{$z$}.}
\end{figure*}
We use the conventions established in~\cite{boer} for the
asymmetries.  Being $T$-odd, $h_1^\perp$ appears with the $H_1^\perp$, 
the $T$-odd fragmentation function in observable quantities.
In particular, the following weighted 
SIDIS cross section projects out a leading double $T$-odd
$\cos2\phi$ asymmetry, 
\bea
{\langle \frac{\vert P^2_{h{\perp}} \vert}{M M_h} \cos2\phi \rangle}
{\scriptscriptstyle_{UU}}&=& 
\frac{\int d^2P_{h\perp} \frac{\vert P^2_{h\perp}\vert}{M M_h}
\cos 2\phi\,  d\sigma}
{\int d^2 P_{h\perp}\, d\sigma} 
\nn &=&\frac{{8(1-y)} \sum_q e^2_q h^{\perp(1)}_1(x) z^2 H^{\perp(1)}_1(z)}
{{(1+{(1-y)}^2)}  \sum_q e^2_q f_1(x) D_1(z)}
\label{ASY} 
\nn
\eea
where the subscript $UU$ indicates unpolarized beam and 
target (Note: The non-vanishing 
$\cos2\phi$ asymmetry originating from $T$-even distribution and 
fragmentation function 
appears only at order $1/Q^2$~\cite{CAHN,kotz0,OBDN}).
\begin{figure*}
\includegraphics[height=6.0 cm]{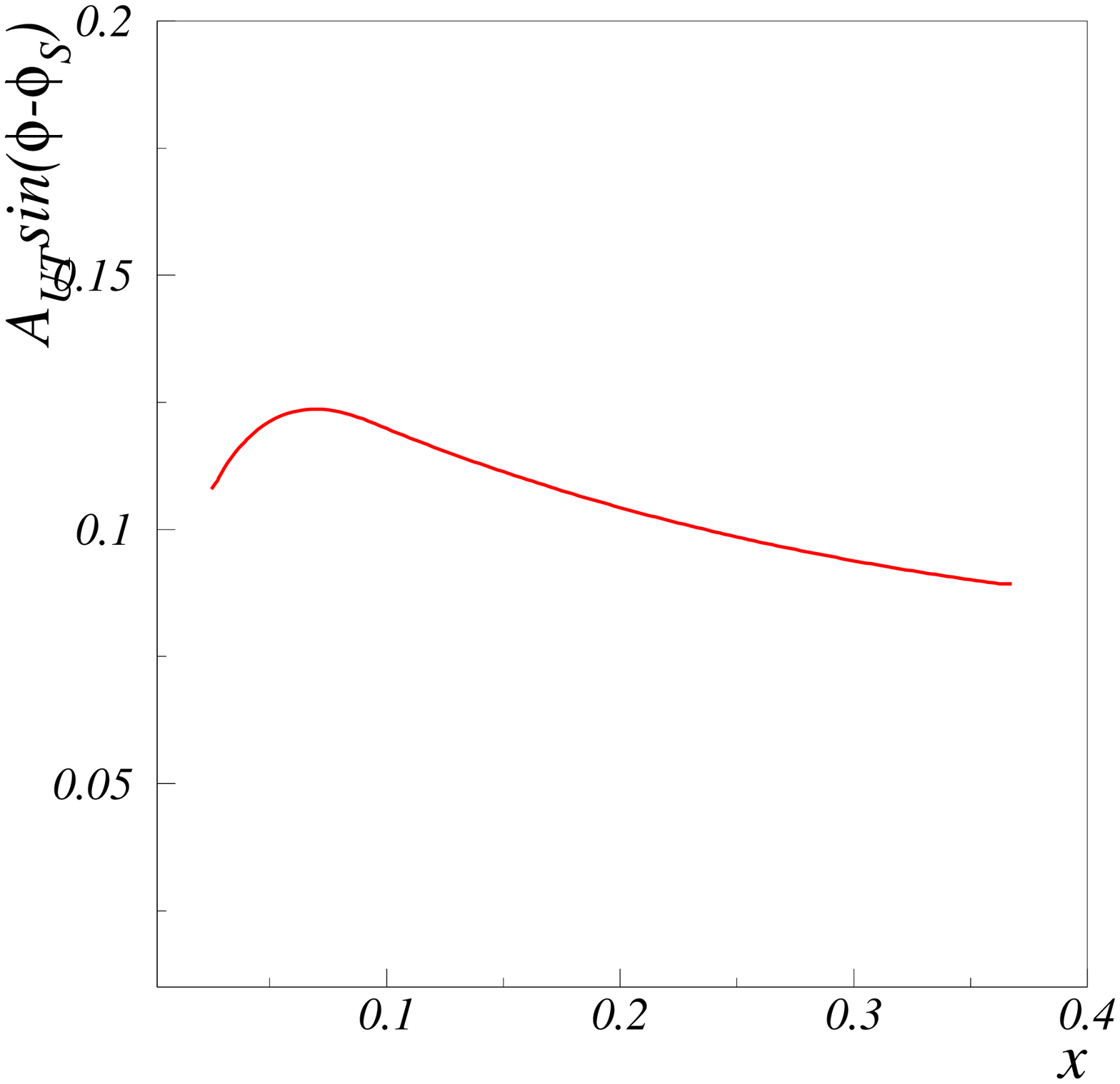}
\includegraphics[height=6.0 cm]{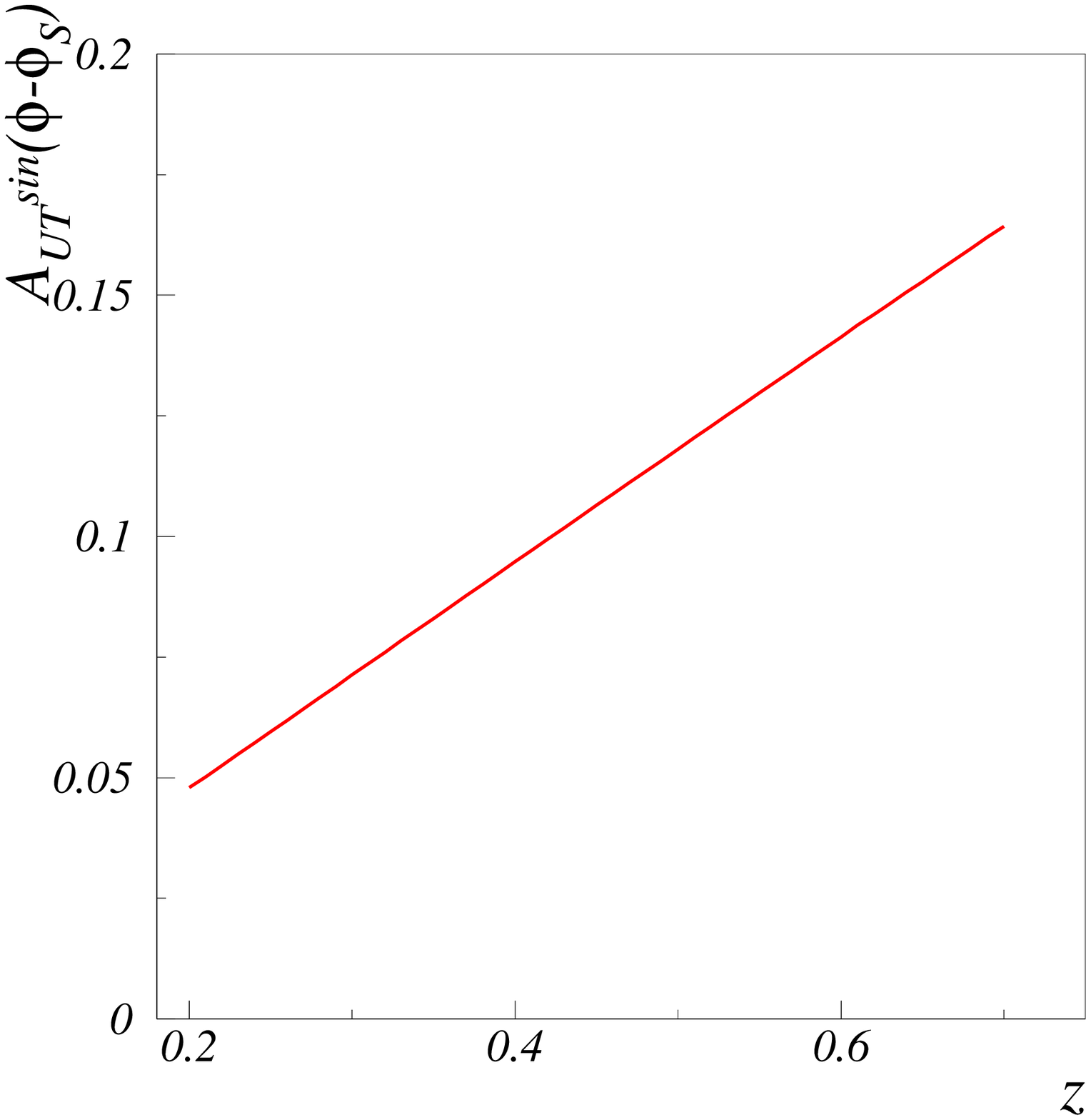}
\caption{\label{sivers}Left Panel: The 
\protect{$A^{\scriptscriptstyle\sin(\phi-\phi_S)}_{\scriptscriptstyle UT}$ 
$x$ }
dependent Sivers asymmetry.
Right Panel: The 
\protect{$A^{\scriptscriptstyle\sin(\phi-\phi_S)}_{\scriptscriptstyle UT}$
 $z$} dependent Sivers asymmetry.}
\includegraphics[height=6cm]{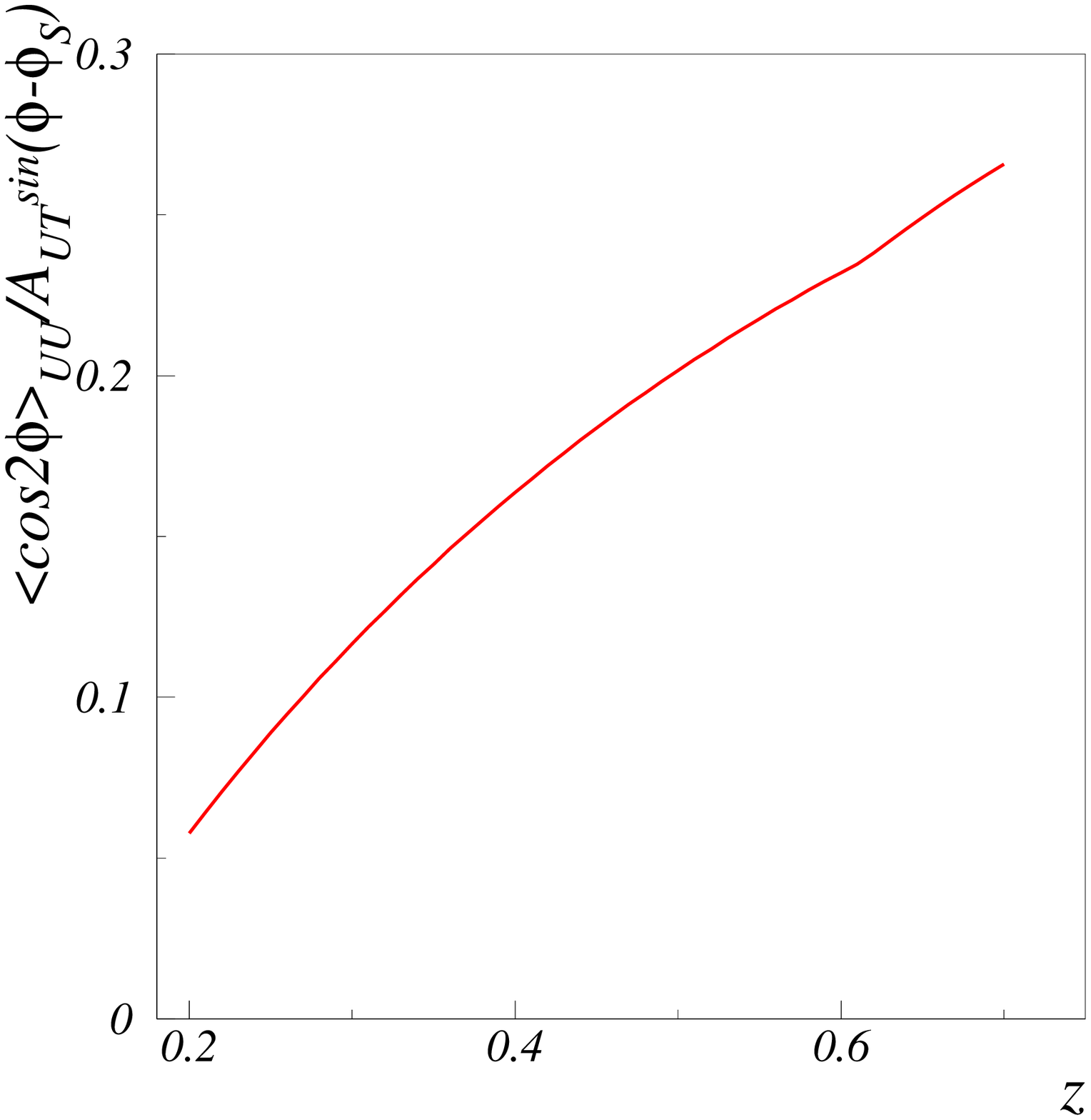}
\caption{\label{ratio} The $z$ dependence of 
the ratio of the ${<\cos 2\phi>}_{\scriptscriptstyle UU}$ 
to  $A^{\scriptscriptstyle\sin(\phi-\phi_S)}_{\scriptscriptstyle UT}$ 
asymmetries.}
\end{figure*}
Additionally, the SSA characterizing the
so-called Sivers effect is
\bea
\langle \frac{\vert P_{h\perp} \vert}{M} 
\sin(\phi-\phi_S) \rangle_{\scriptscriptstyle UT}
&=& 
\frac{\int d^2P_{h\perp} \frac{\vert P_{h\perp}\vert}{M}
\sin (\phi-\phi_S) \, d\sigma}
{\int d^2 P_{h\perp} \, d\sigma} 
 \nn  &&  \hspace{-2cm} = 
\frac{(1+{(1-y)}^2) \sum_q e^2_q f^{\perp(1)}_{1T}(x) z D^q_1(z)}
{{(1+{(1-y)}^2)}  \sum_q e^2_q f_1(x) D_1(z)},
\label{ASY1} 
\nn
\eea
where the subscript $UT$ indicates unpolarized beam and transversely
polarized target.
The functions $h_1^{\perp (1)}(x)$, $f_{1T}^{\perp (1)}(x)$, and 
$H_1^{\perp (1)}(z)$ are the weighted
moments of the distribution and fragmentation functions~\cite{kotz}
where $M$ and $M_h$ are the mass of the target proton and produced hadron
respectively.  We evaluate the ${\langle \cos2\phi \rangle}_{UU}$ and 
$A^{\sin(\phi-\phi_S)}_{UT}$ asymmetries, under the $u$-quark dominance, 
obtained from the  approximations,
\bea
{\langle \cos2\phi \rangle}_{\scriptscriptstyle UU} \approx 
\frac{M M_h}{\langle P^2_{h\perp} \rangle} 
{\langle \frac{\vert P^2_{h\perp} \vert} 
{M M_h} \cos2\phi \rangle}_{\scriptscriptstyle UU}\, ,
\eea
and 
\bea
\A^{\sin(\phi-\phi_S)}_{\scriptscriptstyle UT} \approx 
2 \frac{M}{\langle P_{h\perp} \rangle} 
{\langle \sin(\phi-\phi_S) \rangle}_{\scriptscriptstyle UT} 
\label{EXP1}
\eea
in the  HERMES kinematic range corresponding to, 
$1$ GeV$^2$ $\leq Q^2 \leq 15$ GeV$^2$, $4.5$ GeV 
$\leq E_{\pi} \leq 13.5$ GeV, $0.2 \leq z \leq 0.7$, $0.2 \leq y \leq 0.8$, 
and taking $<P^2_{h\perp}> = 0.25$ GeV$^2$ and  $<P_{h\perp}> = 0.4$ GeV 
as input.  The Collins ansatz~\cite{cnpb93,kotz} for the analyzing power of 
transversely polarized quark fragmentation 
function $H^{\perp}_1(z)$, has been  adopted~\cite{OBDN}. 
For $D_1(z)$, the simple parameterization from Ref.~\cite{REYA} was used. 
In Figs.~\ref{A} and \ref{sivers} the ${\langle \cos2\phi \rangle}_{\scriptscriptstyle UU}$ and 
$A^{\scriptscriptstyle \sin(\phi-\phi_S)}_{\scriptscriptstyle UT}$ of 
Eqs. (\ref{EXP1}) 
for $\pi^+$ production on a proton target is presented 
as a function of $x$ and $z$, respectively. 
Using $\Lambda_{QCD}=0.2$ GeV, and $\mu=0.8$ GeV,
Fig.~\ref{A} indicates approximately a $2\%$ $\cos2\phi$ asymmetry and 
Fig.~\ref{sivers}  a $10\%$  $\sin(\phi-\phi_S)$ asymmetry.
In  Fig.~\ref{ratio} we plotted the ratio of the 
$z$-dependence of the double $T$-odd
$\cos2\phi$ to the single spin $\sin(\phi-\phi_S)$ asymmetry.  The result is 
proportional to the $z$ times
the analyzing power of transversely polarized quark fragmentation
and reflect the equality of $h^\perp_1$ and $f^{\perp}_{1T}$
functions in our approach.

\section{Conclusion}
The interdependence of intrinsic transverse
quark momentum and angular momentum conservation are intimately
tied with studies of transversity. 
This was demonstrated previously  
from analyses of the tensor charge in the context
of the axial-vector dominance approach to exclusive meson 
photo-production~\cite{gamb_gold}, and to SSAs in 
SIDIS~\cite{gold_gamb,gamb_gold_ogan}.  In 
the case of unpolarized beam and target, we have predicted that at HERMES
energies there is a sizable $\cos2\phi$ asymmetry associated with the
asymmetric distributions of transversely polarized quarks inside
unpolarized hadrons.  In order to evaluate this asymmetry we
have modeled the quark intrinsic momentum with a Gaussian 
distribution~\cite{gamb_gold}.  Further we have predicted
the Sivers asymmetry, and as a
a check on this approach, we also predict
the ratio of the $\cos2\phi$ to $A^{sin(\phi-\phi_S)}$
asymmetries. This ratio is consistent 
with the ansatz of Collins~\cite{cnpb93}. 

Beyond these model calculations it is clear that final state interactions
can account for  SSAs.  In addition, it has been shown that other 
mechanisms, ranging from  initial
state interactions to the non-trivial phases  of light-cone wave 
functions~\cite{bhm,bhs} can account for SSAs.
These various mechanisms can be understood from the
context of gauge fixing as it impacts the gauge link operator
in the transverse momentum quark distribution functions~\cite{ji,ji2}.
Thus using rescattering as a mechanism to generate 
$T$-odd distribution functions 
opens a new window into the theory and phenomenology of transversity in 
hard processes.
\begin{acknowledgments} 
We thank Piet Mulders and the organizers of
the ESOP Workshop (December $13^{th}-14^{th}, 2002$)  
where this work was presented. We appreciate  productive discussions
with  S. Brodsky, D. S.  Hwang, D. Boer, F. Yuan and A. Metz. 
LPG is supported in part by funds provided from a
Research Development Grant, Penn State Berks, and GRG from 
the US Department of Energy,   {\small DE-FG02-29ER40702}.
\end{acknowledgments}
\bibliography{apssamp}
\bb{99}

\bibitem{gold_morv}
G. R. Goldstein and M. Moravcsik, Ann.  Phys {\bf 98}, 128 (1976). 

\bibitem{jaffe91}
X.\ Artu and M.\ Mekhfi, Z. Phys. C {\bf 45}, 669 (1990) ;
R.\ L.\ Jaffe and X.\ Ji, Phys. Rev. Lett. {\bf 67}, 552 (1991) ;
Nucl. Phys. {\bf B375}, 527 (1992).

\bibitem{soffer95}
J.\ Soffer, Phys. Rev. Lett. {\bf 74}, 1292 (1995); G.\ R. \ Goldstein,
R.\ L.\ Jaffe and X.\ Ji, Phys. Rev. D {\bf 52}, 5006 (1995).

\bibitem{ralston79}
J. Ralston and D. E. Soper, Nucl. Phys. {\bf B152},109 (1979).

\bit{cnpb93} J.C.~Collins, Nucl. Phys. {\bf B396},161 (1993).

\bibitem{RHIC} G. Bunce, N. Saito, J. Soffer, W. Vogelsang, Ann. Rev. 
Nucl. Part. Sci. {\bf 50}, 525 (2000).

\bibitem{HERMES} A. Airapetian {\it et al.}, Phys. Rev. Lett.
 {\bf 84}, 4047 (2000).

\bibitem{cds} P. Hoodbhoy, Phys.Rev. D {\bf 65}, 077501 (2002);
J. Collins and M. Diehl, Phys. Rev D {\bf 61}, 114015 (2000) ;
M. Diehl, T. Gousset and B. Pire, Phys. Rev. D {\bf 59}, 034023 (1999).

\bibitem{pire} D.Y. Ivanov, B. Pire, L. Szymanowski,
O.V. Teryaev,  Phys.Lett.B {\bf 550}, 65 (2002).

\bibitem{kotz0} A. M. Kotzinian, Nucl. Phys {\bf B441}, 234 (1995).

\bibitem{mulders2} R. D. Tangerman and P. J. Mulders,
Phys. Lett. B {\bf 352}, 129 (1995) ; Phys. Rev. D {\bf 51}, 3357 (1995);
Nucl. Phys. {\bf B461}, 197 (1996).

\bit{ANSL} M. Anselmino and F. Murgia, Phys. Lett B {\bf 442}, 470 (1998);
M. Anselmino, V. Barone, A. Drago and F. Murgia, hep-ph/0209073.

\bibitem{boer} D. Boer and P. J. Mulders, Phys. Rev. D {\bf 57}, 5780 (1998).

\bibitem{cplb} J. C. Collins, Phys. Lett. B {\bf 536}, 43 (2002).

\bibitem{bland} L.C. Bland, hep-ex/0212013; G. Rakness, hep-ex/0211068.
\bibitem{gold_gamb} G. R. Goldstein and L. Gamberg, hep-ph/0209085,
To be published
in the {\em Proceedings of $31^{\rm st}$ International Conference on 
High Energy Physics (ICHEP 2002)}, Amsterdam, The Netherlands, Jul 2002. 

\bibitem{gamb_gold_ogan} L. Gamberg, G. R. Goldstein and 
K.A.~Oganessyan, hep-ph/0211155, To be Published in the {\em Proceedings
of the $15^{\rm th}$ International Spin Physics Symposium (SPIN 2002)}, 
Long Island, New York, September 2002.

\bibitem{bbh} D. Boer, S. Brodsky, D.S. Hwang hep-ph/0211110 .

\bit{sivers} D. Sivers, Phys. Rev D 41 (1990) 83 ; Phys. Rev. D {\bf 43}, 261 
(1991).

\bibitem{gamb_gold} L. Gamberg and G. R. Goldstein, Phys.
Rev. Lett. {\bf 87}, 242001 (2001).

\bibitem{gold_owens} G. R. Goldstein and J. F. Owens,
Phys. Rev. D {\bf 7}, 865  (1973) ; Nucl. Phys. {\bf B71}, 461 (1974).

\bibitem{gold_owens2}
G.R. Goldstein and J.F. Owens, Nucl. Phys. {\bf B103}, 145 (1976).

\bibitem{rodriq} R. Jakob, P.J. Mulders and J. Rodriques, Nucl. Phys.
{\bf A626}, 937 (1997).

\bibitem{bhs}
S. Brodsky, D.S. Hwang and I. Schmidt, Phys. Lett. B {\bf 530}, 99 (2002).

\bibitem{ji} X. Ji and F. Yuan, Phys. Lett. B {\bf 543}, 66 (2002).

\bibitem{ji2}A.V. Belitsky, X. Ji and F. Yuan, hep-ph/0208038.

\bit{col82} J. C. Collins and D. E, Soper, Nucl. Phys. {\bf B194},
445 (1982); J. C. Collins, D. E, Soper and G. Stearman in
{\em Perturbative Quantum Chromodynamics} ed. A. H. Mueller (World
Scientific, 1989), p. 1.

\bibitem{ellis} R.K. Ellis, W.J. Stirling and B.R. Webber, {\it QCD and
Collider Physics} (Cambridge University Press, Cambridge, U.K. 1996),
p.305.
\bibitem{new} The details 
of this calculation will be presented in a longer publication; 
L. Gamberg, G. R. Goldstein and K.A.  Oganessyan, {\em In preparation}.
\bibitem{GRV} M.~Gl\" uck, E.~Reya, and A.~Vogt, Z. Phys. C {\bf 67}, 433 
(1995).

\bibitem{kotz} A. M. Kotzinian and P. J. Mulders,
Phys. Lett. B {\bf 406}, 373 (1997).

\bit{CAHN} R.N. Cahn, Phys. Lett.  B {\bf 78}, 269 (1978) ; 
Phys. Rev. D {\bf 40}, 3107 (1989). 

\bit{OABD} K.A.~Oganessyan,  H. R. Avakian,  N. Bianchi and P. Di Nezza, 
Eur. Phys. J.  C {\bf 5}, (1998).

\bit{OBDN} K.A.~Oganessyan, N. Bianchi, E. De Sanctis, and 
W.D. Nowak, Nucl. Phys. {\b A689}, 784 (2001).

\bit{REYA} E.~Reya, Phys. Rep. {\bf 69}, 195 (1981).  

\bit{bhm} S. J. Brodsky, P. Hoyer, N. Marchal, S. Peign\'{e}, and
F. Sannino, Phys. Rev. D {\bf 65}, 114025 (2002).

\eb
\end{document}